\documentstyle[12pt]{article}
\begin{document}
\begin{center}
PARAMETER MISMATCHES AND PERFECT ANTICIPATING SYNCHRONIZATION IN BI-DIRECTIONALLY COUPLED EXTERNAL CAVITY LASER DIODES\\
E. M. Shahverdiev, S.Sivaprakasam and K. A. Shore \\
School of Informatics, University of Wales, Bangor, Dean Street, Bangor, LL57 1UT, Wales, UK\\
ABSTRACT\\
\end{center}
We study {\it perfect} chaos synchronization between two bi-directionally coupled external cavity semiconductor lasers and demonstrate for the first time that mismatches in laser photon decay rates can explain the experimentally observed anticipating time in synchronization.\\
PACS numbers(s):05.45.Xt,05.45.Vx,42.55.Px,42.65.Sf\\
\indent Chaos synchronization [1] is of fundamental importance in a variety of complex physical, chemical and biological systems [2]. Application of chaos synchronization has been advanced in secure communications, optimization of non-linear system 
performance, modeling brain activity and pattern recognition [2]. Time-delay systems are ubiquitous in nature, technology and society because of finite signal transmission times, switching speeds and memory effects [3]. Therefore the study of chaos 
synchronization in these systems is of considerable practical significance. Because of their ability to generate high-dimensional chaos, time-delay systems are good candidates for secure communications based on chaos synchronization. In this context 
particular emphasis is given to the use of chaotic external cavity semiconductor lasers [4].\\
\indent Following the discovery of anticipating synchronization by Voss [5], Masoller  studied theoretically and numerically   anticipating synchronization in unidirectionally coupled lasers and showed  that the anticipating time should be equal to the difference between round-trip time of the light in the transmitter's external cavity and the time of flight between the lasers [6]. The first experimental observation of anticipating synchronization between two bi-directionally coupled external cavity laser diodes was  reported recently [7].In a bi-directional system such anticipating chaos is rather robust but it has proved to be rather difficult to obtain a reproducible  demonstration of anticipating synchronization in unidirectionally coupled laser diodes. It is noted also that it was found experimentally [7] that the solitary receiver laser anticipates the chaotic transmitter by the time of flight between the lasers. At present there is no full theoretical explanation of the experimental results.\\
\indent In this Brief Report we demonstrate the possibility  of {\it perfect} chaos synchronization between two bi-directionally coupled external cavity semiconductor lasers. We demonstrate, for the first time, that  mismatches in laser photon decay rates can explain the time of flight anticipating time synchronization between the laser diodes. In a recent paper [8] we have demonstrated the role which parameter mismatches plays in the explanation of the coupling-delay lag time synchronization in uni-directionally coupled systems. Knowledge of the exact time shift between the synchronized states is of obvious considerable practical importance for the recovery of message at the receiver of a chaotic communication system [4,9].\\ 
\indent An appropriate framework for treating the evolution of the electric field of external cavity laser diodes is provided by the widely utilised Lang-Kobayashi equations [10]. Suppose that the master laser is described by the equations 
$$\hspace*{0.2cm}\frac{dE_{1}}{dt}=\frac{(1+\imath\alpha_{1})}{2}(\frac{G_{1}(N_{1}-N_{01})}{1+s_{1}\vert E_{1} 
\vert^{2}}-\gamma_{1})E_{1}(t)+k_{1}E_{1}(t-\tau_{1})\exp(-\imath\omega\tau_{1})+k_{3}E_{2}(t-\tau_{2})\exp(-\imath\omega\tau_{2}),$$
$$\hspace*{6cm}\frac{dN_{1}}{dt}=J_{1}-\gamma_{e1} N_{1}-\frac{G_{1}(N_{1}-N_{01})}{1+s_{1}\vert E_{1} \vert^{2}}\vert 
E_{1} \vert^{2},\hspace*{3.4cm}(1)$$
is coupled bidirectionally with the slave laser described by equations 
$$\hspace*{0.2cm}\frac{dE_{2}}{dt}=\frac{(1+\imath\alpha_{2})}{2}(\frac{G_{2}(N_{2}-N_{02})}{1+s_{2}\vert E_{2} 
\vert^{2}}-\gamma_{2})E_{2}(t)+k_{2}E_{2}(t-\tau_{1})\exp(-\imath\omega\tau_{1})+k_{3}E_{1}(t-\tau_{2})\exp(-\imath\omega\tau_{2}),$$
$$\hspace*{6cm}\frac{dN_{2}}{dt}=J_{2}-\gamma_{e2} N_{2}-\frac{G_{2}(N_{2}-N_{02})}{1+s_{2}\vert E_{2} \vert^{2}}\vert 
E_{2} \vert^{2},\hspace*{3.4cm}(2)$$
where $E_{1,2}$ are the slowly varying complex fields for the master and slave lasers,respectively;$N_{1,2}$ are the carrier densities;$\gamma_{1,2}$ are the cavity losses;$\alpha_{1,2}$ are the linewidth enhancement factors;$G_{1,2}$ are the optical 
gains;$k_{1,2}$ are the feedback levels;$k_{3}$ is the coupling rate;$\omega$ is the optical frequency without feedback (no frequency detuning between the two lasers);$\tau_{1}$ is the round-trip time in the external cavity;$\tau_{2}$ is the time of flight between the master laser and the slave laser-coupling delay time;$J_{1,2}$ are the injection currents;$\gamma_{e1,e2}^{-1}$ are the carrier 
lifetimes;$s_{1,2}$ are the gain saturation coefficients.\\
\indent We show that mismatches between the master and slave laser photon decay rates $\gamma_{1}\neq\gamma_{2}$ can result in the  experimentally observed time of flight anticipating synchronization time. Mathematically the intensities of the master and slave lasers should be related by $$\hspace*{5cm}I_{1}=I_{2,\tau_{2}}\hspace*{10.1cm}(3)$$
Throughout this paper $x_{\tau}\equiv x(t-\tau)$. We also assume an analogous synchronization manifold for the carrier densities:$N_{1}=N_{2,\tau_{2}}$. Using eq.(2) we write the dynamical equation for the $E_{2,\tau_{2}}$ in the following manner:
$$\hspace*{-1.7cm}\frac{dE_{2,\tau_{2}}}{dt}=\frac{(1+\imath\alpha_{2})}{2}(\frac{G_{2}(N_{2,\tau_{2}}-N_{02})}{1+s_{2}\vert E_{2,\tau_{2}} \vert^{2}}-\gamma_{2})E_{2,\tau_{2}}+k_{3}E_{1,2\tau_{2}}\exp(-\imath\omega\tau_{2}),$$
In accordance with experiments [7] we consider the case of a solitary slave laser,i.e. $k_{2}=0$. It is  assumed that, except for the photon decay rates, the laser parameters are identical. Then we find that the equations for $E_{1}$ and $E_{2,\tau_{2}}$ will be identical and therefore perfect synchronization (3) will be possible if conditions
$$\hspace*{6cm}\frac{(1+\imath\alpha)}{2}\gamma_{1}=\frac{(1+\imath\alpha)}{2}\gamma_{2}\mp k_{3}\exp(-\imath\omega\tau_{2}),\hspace*{3.3cm}(4)$$ 
(where $\alpha=\alpha_{1}=\alpha_{2}$)\\
and 
$$\hspace*{4cm}k_{1}E_{1}(t-\tau_{1})\exp(-\imath\omega\tau_{1})=k_{3}E_{1}(t-2\tau_{2})\exp(-\imath\omega\tau_{2}).\hspace*{3.3cm}(5)$$
are met.\\
One can easily re-write condition (4) in the  more appealing form:
$$\hspace*{6cm}(\gamma_{1}-\gamma_{2})^{2}=\frac{4k_{3}^{2}}{1+\alpha^{2}},\hspace*{7.2cm}(6)$$
At first glance it may seems rather unusual to impose  conditions on the chaotic transmitter itself as in (5). But, given the presence of six free parameters, this condition is not particulary restrictive.\\
(Indeed ,in certain cases no such restrictions are needed at all. For example, assuming $\tau_{1}=2\tau_{2}$ one finds that the perfect synchronization manifold (3) exists  for $k_{1}=k_{3}$ and $\omega\tau_{2}=2\pi n$(where $n=0,1,2...$). We also notice that the synchronization 
manifolds $I_{1}=I_{2,\tau_{2}}$ and  $I_{1}=I_{2,\tau_{1}-\tau_{2}}$ 
are identical for $\tau_{1}=2\tau_{2}$, but in general, $I_{1}=I_{2,\tau_{1}-\tau_{2}}$ is not the synchronization manifold.) Our approach, in principle allows also for anticipating synchronization with time of flight anticipation time, even in the case 
when $\tau_{1}=\tau_{2}$, as  was reported  experimentally [7]. Thus we conjecture that mismatches in laser cavity decay rates- as could be expected in practice- renders practically impossible the observation of anticipating synchronization between unidirectionally coupled lasers. Here we would like to underline that   mismatches in laser cavity decay rates are perfectly adequate to  explain the time of flight lag time synchronization in uni-directionally coupled chaotic external cavity laser diodes [11]. In [11] it is shown that the perfect lag synchronization manifold $I_{2}=I_{1,\tau_{2}}$ exists simply under the conditions (4) and $k_{1}=k_{2}$.\\
\indent We conclude this Brief Report with the following remarks. Usually parameter mismatches are considered to have a detrimental effect on the synchronization quality between coupled identical systems: in the case of small parameter mismatches 
the synchronization error does not decay to zero with time, but can show small fluctuations about zero or even a non-zero mean value [6]. Larger values of parameter mismatches can even result in the loss of synchronization. However it appears that in reality the relation between chaos synchronization in time-delayed systems and parameter mismatches is quite intricate and complex. In a recent paper [8] we have shown that parameter mismatches can change the time shift between the synchronized states;moreover we have presented an example where the presence of parameter mismatches is the {\it only way to achieve chaos synchronization} between two unidirectionally coupled time-delayed systems. In the present Brief Report we have shown that perfect anticipating synchronization between two bi-directionally coupled external cavity laser diodes is possible in the presence of parameter mismatches. As knowledge of the 
time shift between the synchronized states is of considerable practical importance for the message recovery in communications and information processing using chaos control methods, further research on relation between chaos synchronization and parameter mismatches would be desirable. \\
This work is supported by UK EPSRC under grants GR/R22568/01 and GR/N63093/01.

\end{document}